# Progress in Fission Fragment Rocket Engine Development and Alpha Particle Detection in High Magnetic Fields

Sandeep Puri*[1], Cuikun Lin[1], Andrew Gillespie[1], Ian Jones[1], Christopher Carty[1], Mitchell Kelley[1], Ryan Weed*[2], Robert V. Duncan*[1]

[1]Center for Emerging Energy Sciences, Department of Physics and Astronomy, Texas Tech University, Lubbock, Texas, USA

[2]Positron Dynamics, Livermore, CA

**Corresponding Authors:** sandpuri@ttu.edu; robert.duncan@ttu.edu; ryan.weed@positrondynamics.com.

**Abstract**

In this article, we present our recent experiments on fission fragment rocket propulsion, and on an innovative new design for an alpha particle detection system that has been inspired by these rocketry results. Our test platform, which operates within high magnetic fields 3 T over a large cross-section (~ 40 cm diameter), has been used as a test platform to evaluate the containment and thrust within a future fission-fragment rocket engine (FFRE). This much more efficient nuclear rocket propulsion FFRE design was first proposed in the 1980s with the intent of greatly reducing transit times in long-duration space travel. Our objective is to enhance the operational efficiency of this nuclear rocket while gaining deeper insights into the behavior of fuel particles and of the fission-fragment ejecta within strong magnetic fields experimentally. Through a combination of simulations and experimental work, we established a method for the production and detection of alpha particles as a surrogate for fission fragments. To achieve this, we employed Americium-241 ($^{241}$Am) sources, which were situated within a cylindrical vacuum chamber positioned in a 3-T Siemens MRI superconducting magnet. By simulating, measuring, and analyzing the emitted alpha particle flux, we gained valuable information about the distribution and likelihood of escape of fission fragments in a future FFRE design. This approach could potentially achieve both high specific impulse and power density in advanced nuclear propulsion systems, such as the FFRE. More generally, this work provides a powerful new approach for analyzing ion flux and nuclear particle / nuclear reaction fragments from a wide variety of experimental designs.

## 1.0 Introduction

A human spaceflight mission is a demanding and dangerous venture that presents serious health hazards to the crew.[1], [2] Factors such as ionizing space radiation, altered gravity extremes, and long-term isolation and confinement are among the health hazards faced by the astronauts.[1],

[2] Minimizing these risks is essential in the quest to achieve a successful completion of the mission. One of the solutions to alleviate such challenges would be to improve the efficiency and speed of travel, thereby reducing the mission's duration. To achieve this goal, a rocket with both of the competing capabilities of high thrust, which is needed to escape the gravitational pull near planets, and high impulse, which is needed to achieve high travel speeds between planets, are required.[3] Nuclear rocket propulsion has the potential to provide higher thrust and efficiency than conventional chemical rockets,[4] allowing faster and more efficient travel to Mars or other distant planets and facilitate the delivery of scientific instruments or materials for use in space throughout and beyond our Solar System.[5]

**1.1 Introduction to fission-fragment rocket propulsion**

A number of designs have been proposed to create high-performance nuclear propulsion systems.[4], [6], [7], [8], [9] The majority of the nuclear rockets, like all commercial nuclear reactors today, are designed to deposit the kinetic energy of their fission fragments as heat in the nuclear fuel. This mandates a thermal conversion process to accelerate the rocket's propellent which is usually hydrogen.[9] In 1988, the Idaho National Engineering Laboratory and Lawrence Livermore National Laboratory proposed a new reactor concept.[10] By proposing magnetic fields to guide these fission fragments out of an ultra-low-density core, this proposed system achieved remarkably high exhaust velocities, resulting in specific impulses of approximately one million seconds, all while maintaining significant thrust levels. This design featured a configuration where fuel was strategically placed on the surface of numerous ultra-thin carbon fibers, arranged radially like wheels. These "wheels" typically remained in a sub-critical state. George Chapline at LLNL proposed the use of a fission-fragment rocket powered by $^{242m}$Am, based on the direct heating of a propellant gas by fission fragments generated from fissile materials.[10], [11], [12] In 2000, Ronen *et al*. demonstrated that $^{242m}$Am could maintain sustained nuclear fission as an extremely thin metallic film, measuring less than 1/1000 of a millimeter in thickness. Notably, $^{242m}$Am required only 1% of the mass of $^{235}$U or $^{239}$Pu to reach its critical state. Fast forward to 2005, when Rodney L. Clark and Robert B. Sheldon introduced a new design.[13] Their concept involved nanoparticles of fissionable fuel (or even naturally radioactive decayable fuel) housed within a vacuum chamber subjected to both an axial magnetic field (acting as a magnetic mirror) and an external electric field. As fission occurred, the nanoparticles ionized, causing the resulting dust to become suspended within the chamber. Two significant challenges need to be addressed: firstly, the containment of substantial quantities of fuels at sub-micron scales; and secondly, the establishment of a thrust axis from these isotropically emitted particles. To solve the challenges, we propose a design and demonstrate a model for an advanced nuclear propulsion system, called the fission fragment rocket engine (FFRE).[5], [14] The FFRE achieves high specific impulse and power density by utilizing a fissile fuel particle embedded in an ultra-low-density aerogel matrix,[14], [15], [16], [17] and additional research is required to determine if this fuel assembly may ever be brought into a critical mass assembly. High field, high-temperature superconducting magnets will be used to constrain fission fragment trajectories. Various alpha emitters can be used in the simulation and design of this magnetic thrust approach, as displayed in Table 1 of the supplemental materials.[18] One naturally occurring isotope, thorium-232 ($^{232}$Th), has a half-life of 14 billion years and mainly decays through alpha decay to radium-228 with a decay energy of 4.0816 MeV. Thorium is used to make ceramics, welding rods, camera and telescope lenses, fire brick, heat-resistant paint, metals used in the aerospace industry, and can also be used as a fertile, not fissile, fuel for generating nuclear energy. According to EPA, since thorium is naturally present in the environment, it poses little health hazard,[19] providing great advantages for initial testing, and for possible future rocket launches. Note that thorium is a convenient radioisotope for alpha-decay testing, but it is not capable of undergoing fission on its own, making it unsuitable for direct use as a fuel in a critical reactor design. Nevertheless, it is considered "fertile", as it can absorb a neutron and transform into uranium-233 ($^{233}$U) following a ~ 30-day decay chain, resulting in an outstanding fissile fuel material. $^{241}$Am stands as another isotopic element of significant interest, with a half-life of 432.2 years. $^{241}$Am is the most common isotope of americium as well as the most prevalent isotope of americium in nuclear waste. It is frequently used in ionization-type smoke detectors and has the potential to serve as fuel for long-lasting radioisotope thermoelectric generators. It's worth noting that the health risks associated with both $^{232}$Th and $^{241}$Am are primarily limited to cases of ingestion or inhalation due to its low alpha radiation penetration capability. Furthermore, through a neutron capture reaction, $^{241}$Am can be converted into the isomer $^{242m}$Am, which features a relatively high thermal cross section for fission of 85.7±6.3 barns [20], [21] and one of the lowest known critical masses of any fissile isotope.

**1.2 Introduction to directed thrust experiments**

The Lorentz Force comprises the entire electromagnetic force, ***F***, on the charged particle, and is expressed in **equation 1** as

$$\boldsymbol{F} = q\boldsymbol{E} + q(\boldsymbol{v} \times \boldsymbol{B}) \qquad (1)$$

where ***E*** is the electric field, ***B*** is the magnetic field, and ***v*** is the velocity of the charged particle, and bold-faced symbols are vector quantities. If a charged particle is placed into a region with a strong magnetic field along only one axis and no externally applied electric field, then the longitudinal velocity, $v_L$, along the external magnetic field, in the very low charged particle density limit, will remain constant along the field axis, and the transverse component of the charged particle's velocity $v_T$, orbits about the magnetic field direction with a Larmor Radius, $r$, that is given by

$$r = mv_T/qB \qquad (2)$$

The ability of a charge particle to avoid deflection by a magnetic field is referred to as its 'rigidity' R, often quoted in units of tesla-meters (Tm). The resulting deflection of the charged particle that is moving transverse to the externally imposed magnetic field then has a circular radius r = R/B. For the experiment involving 4 MeV alpha particles (R ~ 0.3 Tm) produced from $^{232}$Th, we can use **equation 2** to estimate the circular motion radius at different magnetic field strengths. If using a 1-T magnetic field, the circular motion radius is approximately 30 cm, whereas when using a 3-T magnetic field, it reduces to about 10 cm. For comparison, the rigidity of the more energetic alpha particles from $^{241}$Am is approximately 0.33 Tm, and the rigidity of the fission fragments of $^{235}$U fission are about twice as great, varying from 0.60 to 0.66 Tm.

Each charged particle will also experience the forces associated with the electric and magnetic fields produced from the close-proximity and motion of all other charged particles in this containment. Hence, as the density of charged particles that are contained within this region of constant vector magnetic field increases, then the interactions of the charged particles with each other will produce more complex motion than the simple Larmor motion described above. One of the challenges in this effort is to determine how large this charge particle density (which is proportional to the rocket thrust) may be without creating thrust loss from this instability resulting in wall collisions before the charged particles exit the spacecraft.

**Equation. 2** is crucial in determining the appropriate vacuum vessel diameter and the bore size for the magnet required for our experiment. Another critical factor to consider is the uniformity of the magnetic field. Taking all these considerations into account, we find that the 3-T superconducting magnet from a commercial MRI imager is very well-suited for this task. For our experiment, we utilize the Siemens 3T MRI machine within the Texas Tech Neuroimaging Institute, as displayed in Figures S1 (a) and (b) of the supplementary materials.[22] This machine offers exceptional magnetic resonance imaging and was carefully designed to maintain a uniform magnetic field of 3 T within the bore. This provided significant advantages for the experiment and drove the requirements for the vacuum chamber design.

**1.3 Introduction to methods of alpha particle detection and imaging**

There are several well-developed technologies for alpha particle detection, including Geiger-Müller tubes,[23] scintillators with photomultipliers,[24] air ionization chambers,[25] spark chambers,[26] and solid-state electronic detectors.[27] Most of these technologies cannot function properly within a large magnetic field because strong magnets can damage electronic components or distort the particle tracks within the air ionization chamber or spark chamber. Faradzhaev, *et al.*, investigated the performance and material selection of magnetic shields for Photomultiplier Tubes (PMTs).[28] Their findings revealed that the PMT's performance is significantly influenced by magnetic fields, even very weak ones such as the geomagnetic field of the Earth, which typically measures ~ $3 \times 10^{-5}$ T. Although a typical permalloy magnetic shield, such as the one utilized within Hamamatsu PMTs, can attenuate this external magnetic field by 0.1% to 0.01%, the PMT performance can still be affected by residual magnetic fields.[28], [29] Additionally, using arrays of PMTs to study the spatial distribution of alpha particles that escape the vacuum chamber to strike the endplate presents significant design and fabrication challenges. The presence of a magnetic field does not alter the trajectory of light, so an effective solution to address this challenge and investigate the spatial distribution of alpha particles that reach the endplate of the vacuum chamber is to create a new alpha particle detection system, ensuring accurate analysis while mitigating the impact of the magnetic field on the optical imaging system, as described in the next section of this report.

The primary focus of this research was to gain insight into the collective behavior of the ejecta particles within the high magnetic field of our 3-T system. To accomplish this, we devised and constructed an innovative alpha particle detection system. Additionally, we conducted simulations using COMSOL Multiphysics 6.2 software for Multiphysics Simulation. For our preliminary investigations, we selected $^{232}$Th and $^{241}$Am as the fuel material that provided the alpha particles for our surrogate fission-fragment testing. This compound was positioned within a cylindrical vacuum chamber, and then placed inside the Siemens 3 T MRI superconducting magnet. Through this design, simulations and experiments were performed to carefully track the trajectories of emitted alpha particles. Our aim was to gain a comprehensive understanding of the kinetic energy distribution and escape probability under this magnetic field as a function of the alpha particle density.

## 2.0 Experiment and Discussions

### 2.1 Range of alpha particles in air

The range of energetic alpha particles is determined by the stopping power of the material that they pass through. The range is defined as the distance in a medium at which the alpha particle loses most of its energy due to collisions with electrons.[30] This distance depends on the particle's type and energy as well as the material it travels through. Fenyves and Haiman found that the range of alpha particles for energies between 4-7 MeV can be calculated using **equation 3**.[31] For example, a 5.5 MeV alpha particle emitted by $^{241}$Am can travel approximately 3.9 cm in air, whereas the 4.0 MeV alpha particle from $^{232}$Th has a range of about 2.4 cm. This short stopping distance is due to the double-positive charge, the mass, and the speed of alpha particles, which results in their immense capability to ionize air.

$$R_{air}[\text{cm}] = 0.3\, E\, [MeV]^{3/2} \qquad (3)$$

where $R_{air}$ is the stopping distance in units of centimeters and $E$ is the particle energy in units of MeV. This value for 4 MeV alpha particles is quite similar to the one found from calculations using the open-source software called 'Stopping and Range of Ions in Matter' (SRIM),[32] which is 2.53 cm. As the pressure decreases, the range of energetic particles increases as 1/P. The range of energetic alpha particles as a function of pressure, as calculated using SRIM, as displayed in **Table 2** of the supplemental materials.

### 2.2 Vacuum system design

The vacuum chamber and all interior and exterior components were built from strictly non-magnetic materials for mechanical stability in the 3-T field. Materials such as aluminum, stainless steel, gold, silver, and copper may be used. Aluminum, like lithium and magnesium, possesses a crystal structure that renders it non-magnetic. Aluminum forms a stable, tough oxide on its surface that makes it able to withstand harsh environments, and it is strong and light weight, making it ideal for this experimental application.

The design of the aluminum vacuum chamber is illustrated in **Figure 1**, along with a picture of the final product. The vacuum chamber is a seam-welded aluminum pipe (1) of 60" (152 cm) length with inner and outer diameters 13.5" (34.3 cm) and 14" (35.6 cm) respectively. The flanges at both ends are also made of aluminum with a thickness of ½" (1.27 cm) with inner and outer diameters 13.5" (34.3 cm) and 16.0" (40.6 cm), respectively. One end of the pipe is covered by a ¼"-thick aluminum plate with four clearance holes that compress an O-ring seal, and the other end is covered by a 1"- thick clear acrylic window.  The vacuum chamber was tested using a Pfeiffer ASM 340 helium leak detector, and no leak was found within its detection limit.

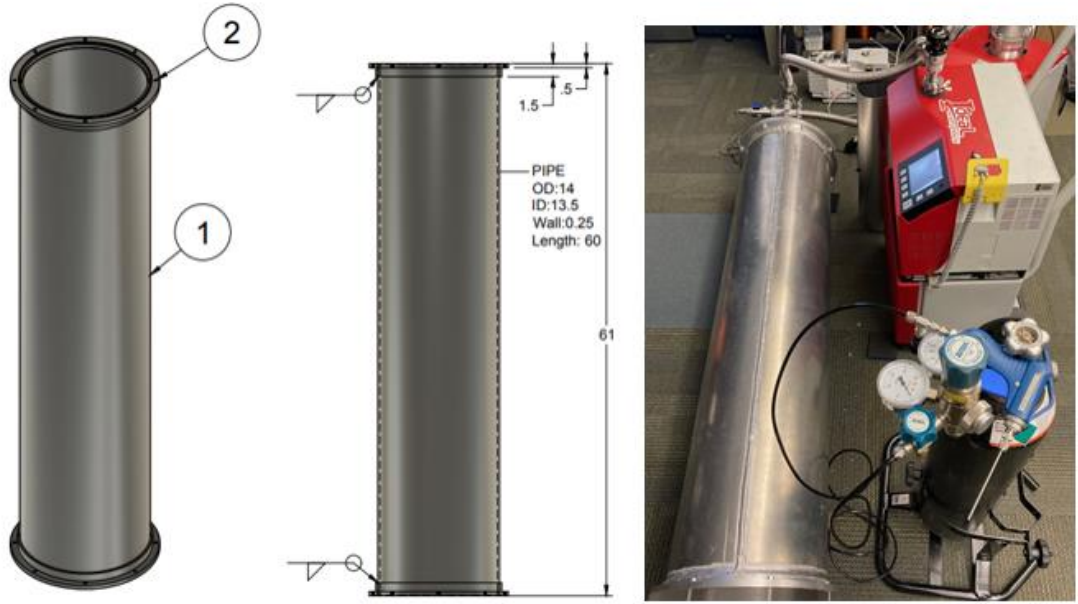

**Figure 1:** Design of the hermetic aluminum vacuum chamber, along with a picture of the final product during leak testing.

**2.3 Alpha particle detection system**

It is important to highlight that the presence of a magnetic field does not alter the trajectory of light. Hence, an effective way to detect the spatial distribution of alpha particles within the magnetic field as they strike the endplate is to design an extended chamber to collect the very low-level intensity light from the alpha particle scintillation on the interior surface of the plastic endplate. This chamber utilizes an ultra-sensitive camera further away from the high magnetic field, ensuring an accurate analysis by mitigating the effects of the magnetic field on the camera. According to Siemens (the manufacturer of the 3-T MRI magnet) the magnetic field is ~ 200 mT at one meter away from the entry port, and only 10 mT at two meters away. As shown in **Figure 2**, the alpha detection chamber is designed to be comprised of several components, including a scintillator material (3), a clear acrylic window that serves as the vacuum chamber endplate (4), a dark chamber (5), a lens, and a detector. For the detector, there are quite a few options, such as a Complementary Metal-Oxide Semiconductor (CMOS) camera, a Single-Photon Avalanche Diode (SPAD) camera, as well as a Charge-Coupled Device (CCD) camera. The science community has shown great interest in the recent developments of SPAD arrays and Multi-Pixel Photon Counters (MPPC) coupled with Silicon Photomultiplier (SiPM) arrays.[33], [34], [35] These devices have transformative applications in Quantum Technology (such as in Quantum Key Distribution), Solid State Flash Light Detection And Ranging (SSF LiDAR), brain activity monitoring using Positron Emission Tomography (PET) medical imaging, etc.[36], [37] However, the development of megapixel SPAD cameras is challenging. A joint project between Swiss research institute EPFL (École Polytechnique Fédérale de Lausanne) and Canon has developed a megapixel SPAD image sensor, but it is not commercially available yet.[38] On the other hand, A CMOS Camera is the economical alternative to SPAD camera, with only slightly less sensitivity. For the scintillator, silver-activated zinc sulfide is an inorganic scintillator material that has been used for many years.

It has a high scintillation efficiency comparable to that of Sodium Iodide Scintillation Detectors (NaI(Tl)), with a theoretical light yield of approximately 100,000 Photons/MeV. However, manufacturers of ZnS(Ag) typically report a lower light yield of about 50,000 photons/MeV.[39] For this alpha detection chamber, we utilized EJ-440 from Eljen Technology, TX as our scintillator material. This material consists of a ZnS(Ag) phosphor that is applied to a clear polyester plastic sheet, making it quite flexible and easy to cut and pattern. The alpha sensitivity of each batch is uniform within a ± 1.5% range.[40] When 10,000 alpha particles with 4 MeV energy are detected, approximately 2 billion photons are produced. Given this, a cooled CMOS camera is a suitable choice for this application. After careful consideration, we have selected the FL-20BW low noise CMOS camera from Tucsen[41] for our experiments. This camera model features a dark current of 0.001 e-/pixel/s and a noise level of 0.6 e-/pixel/s.

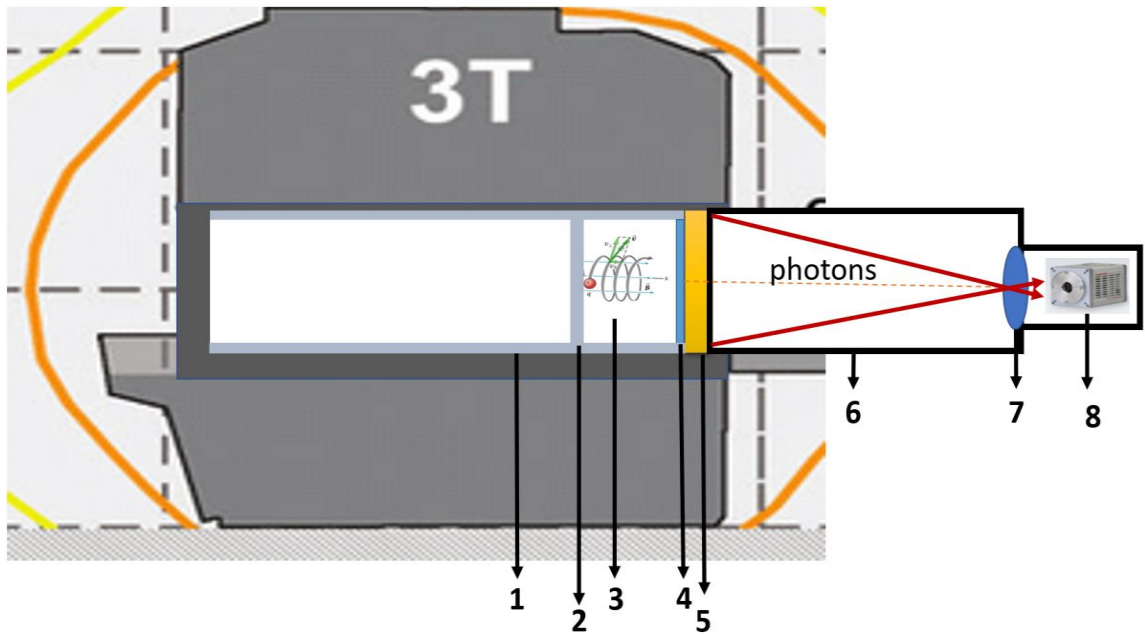

**Figure 2:** An add-on alpha particle detection chamber has been developed, consisting of several components, including a source panel (2) a scintillator material (4), a clear acrylic window (5), a dark chamber (6), a lens (7), and a detector (8). The alpha particles in Larmor orbits (3) within the vacuum chamber (1) are displayed as well.

### 3.0 COMSOL Simulations

We conducted COMSOL simulations to gain a deeper understanding of the behavior of alpha particles and to aid in the design of a vacuum chamber within a 3-T superconducting magnet, and to better understand our experimental results. For our research, we utilized COMSOL Multiphysics 6.1 (COMSOL 6.1) software to design and simulate our proposed model. COMSOL Multiphysics is a physics simulation software program that provides various physical solvers and enables the coupling of multiple physics interface modules such as electrodynamics, acoustics, fluid flow, heat transfer, etc. We can construct models by defining relevant physical quantities (such as material properties, loads, constraints, sources, and fluxes) instead of specifying basic equations. These variables, expressions, or numbers can be directly applied to solids and fluid domains, boundaries, edges, and points, regardless of the computational grid. The software then internally formulates a set of equations representing the entire model. Since our aim was to investigate the behavior of charged particles and their trajectories under the influence of electromagnetic forces, we utilized the Charged Particle Tracing interface, available under the AC/DC > Particle Tracing branch when adding a physics interface. This interface is used to model charged particle orbits under the influence of electromagnetic forces in COMSOL Multiphysics 6.1 for simulating our model. Further details of the simulation are provided in the supporting information.

## 4.0 Alpha particle detection

For the initial tests, we incorporated a 12" EJ-440 scintillator and a 12" ThO$_2$ plate within the aluminium vacuum chamber as depicted in the **Figure S2** in the supplementary information. Following the successful detection of alpha particles, we proceeded to transition to $^{241}$Am, which emits alpha particles with an energy of 5.5 MeV and exhibits higher radioactivity. As seen in **Figure 3(a)** and **3(b)**, four-point alpha emitting $^{241}$Am sources were placed on an aluminium plate, with the radio activities of each $^{241}$Am source labelled in **Figure 3(a)** in units of microcuries. **Figure 3(c)** shows an image captured by the FL-20BW CMOS camera in response to alpha particle exposure with an integration time of 15 minutes with no externally applied magnetic field. The scintillator response is primarily from alpha particles emitted from a sample of $^{241}$Am hitting a scintillator screen, which agrees well with COMSOL simulation. As alpha particles emit isotropically, many particles collide with the wall, and only the alpha particles with a line-or-sight of the scintillator were able to contribute to this diffuse image.

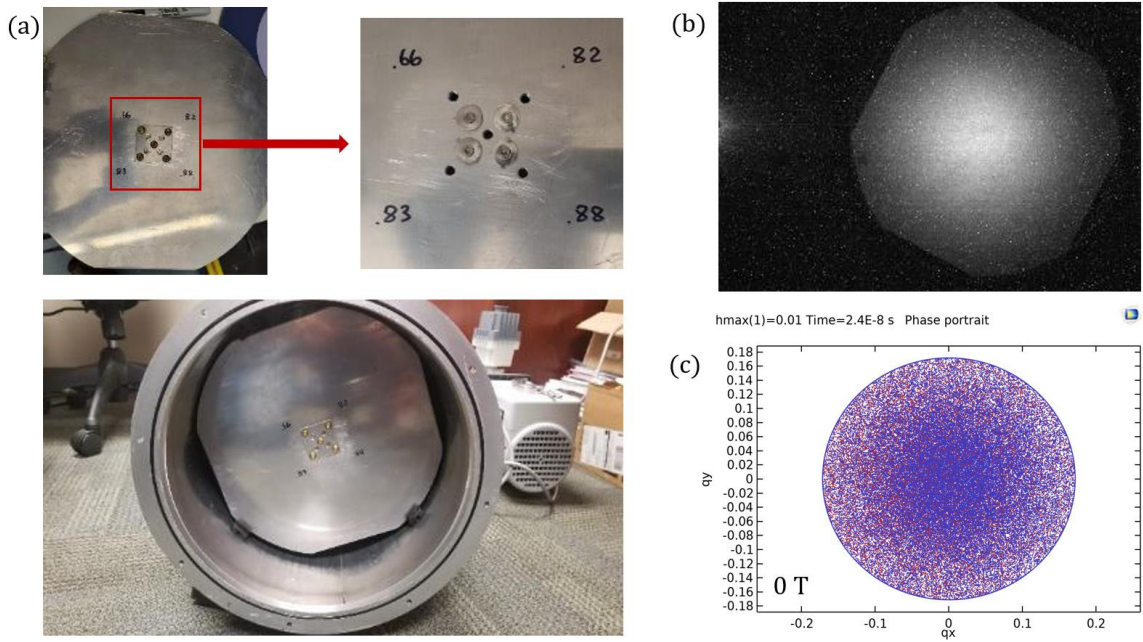

**Figure 3:** (a) *Top:* An experimental set up using 4-point alpha emitting $^{241}$Am source (3.19 µCi total) placed on an alluminum plate, *Bottom:* source panel positioned within the chamber. (b) Experimental results performed at 0 T magnetic field with 15 min exposure time. (c) Phase portrait showing particle positions obtained from the COMSOL simulations for particles released from the 4-point

sources, similar to the four $^{241}$Am point sources in the experiment, at 0 T magnetic field. Here the blue color indicates alpha particles that originated from the left two sources in Figure 3a, and the red color indicates alpha particles that were emitted from the two right sources. The interpenetration of blue and red indicates that the sources were isotropic in their alpha particle emissions, as expected.

When introducing the chamber into the 3-T magnetic field in the MRI room, several challenges arise. As depicted in **Figures S1** and **S4** in the supplementary materials, the experiment preparation process to insert the aluminium apparatus into the 3 T field is crucial. It is essential not to place any magnetic materials, including screws, Kleinflansch (KF) flanges, KF clamps, magnetic vacuum pipes, etc. into the high magnetic field. To address this issue, we replaced all the KF flange components with non-magnetic materials to ensure compatibility with the magnetic field. Additionally, in order for the camera to capture the weak luminescence resulting from alpha particle strikes, the room must be completely dark. To achieve this, we utilized 'Blackout Fabric' from Thorlabs during the experiments to prevent any light leakage into the camera chamber. The aluminium chamber must be evacuated, but the vacuum pump may not be placed into the high field region of the MRI room. Hence a 13-meter-long copper pipe was used to connect the vacuum vessel to a pump located outside of the MRI room, which adequately maintained the vacuum level within the chamber during the alpha particle imaging. The magnetic field at the location of the camera was measured, and it always remained below 10 mT. In **Figure 4 (a)**, the FL-20BW CMOS camera captures an image with the $^{241}$Am panel positioned 20 inches from the scintillator panel within a 3-T magnetic field. Evidently, there is clear magnetic confinement of the alpha particles. Surprisingly, four sources are also observed, forming four distinct circles on the scintillator panel. The successful confinement of the alpha particles allows for direct observation of the image, requiring only a 120-second exposure time. When the $^{241}$Am panel is positioned closer to the scintillator panel at distances of 15" and 10", the luminescence in the images captured by the FL-20BW camera is scarcely observed (**Figure (b-c)**). Consequently, it was necessary to extend the exposure time to 450 seconds for the 15" location (**Figure (e)**) and 900 seconds for the 10" location (**Figure (b)**). Additionally, we observed that, decreasing the distance between the source and the scintillator panel resulted in larger confinement area compared to when the distance was 20". The colors shown in the simulation results in **Figure 5** correspond to alpha particles from the same source placements as described in the Figure 3 caption.

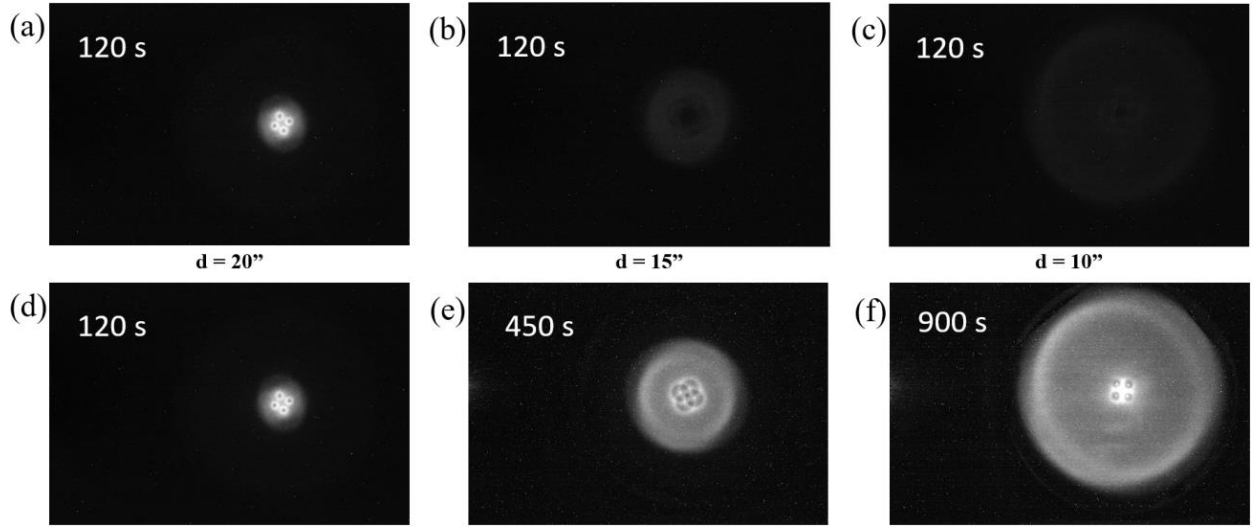

**Figure 4:** Images of the particles hitting a scintillator panel as captured by the FL-20BW CMOS camera. (a — c) Luminescence observed on the scintillator when $^{241}$Am panel was positioned at distances 10", 15" and 20" respectively from the scintillator panel for exposure time of 120-second. Luminescence is scarce in the scintillator when the source panels were positioned at 10" and 15". (d — f) Brighter luminescence observed when exposure time was increased to (e) 450 s for the panel positioned at 15" and (f) 900 s for the panel positioned at 10" from the scintillator panel.

We utilized COMSOL (as described above) to simulate the effects of the electric and magnetic fields that are generated by the ions, in this case alpha particles, within the vacuum can. Despite the inclusion of the electrostatic field in the simulation, the low beam current results in negligible divergence caused by self-potential. Through multiple iterations, the model converges to a self-consistent solution for both the trajectories of alpha particles and the beam potential. **Figure 5** displays these trajectories at positions 20", 15", and 10", with colors indicating the radial displacement of each particle from its initial position. Additionally, we observed confinement effects due to the externally imposed magnetic field, which align well with experimental findings. Another way to visualize particle positions is using the Phase Portrait plot within COMSOL. A Phase Portrait is a 2D plot in which the axes can be arbitrary expressions in terms of the particle position, velocity, or any other variable that can be evaluated for the particles.

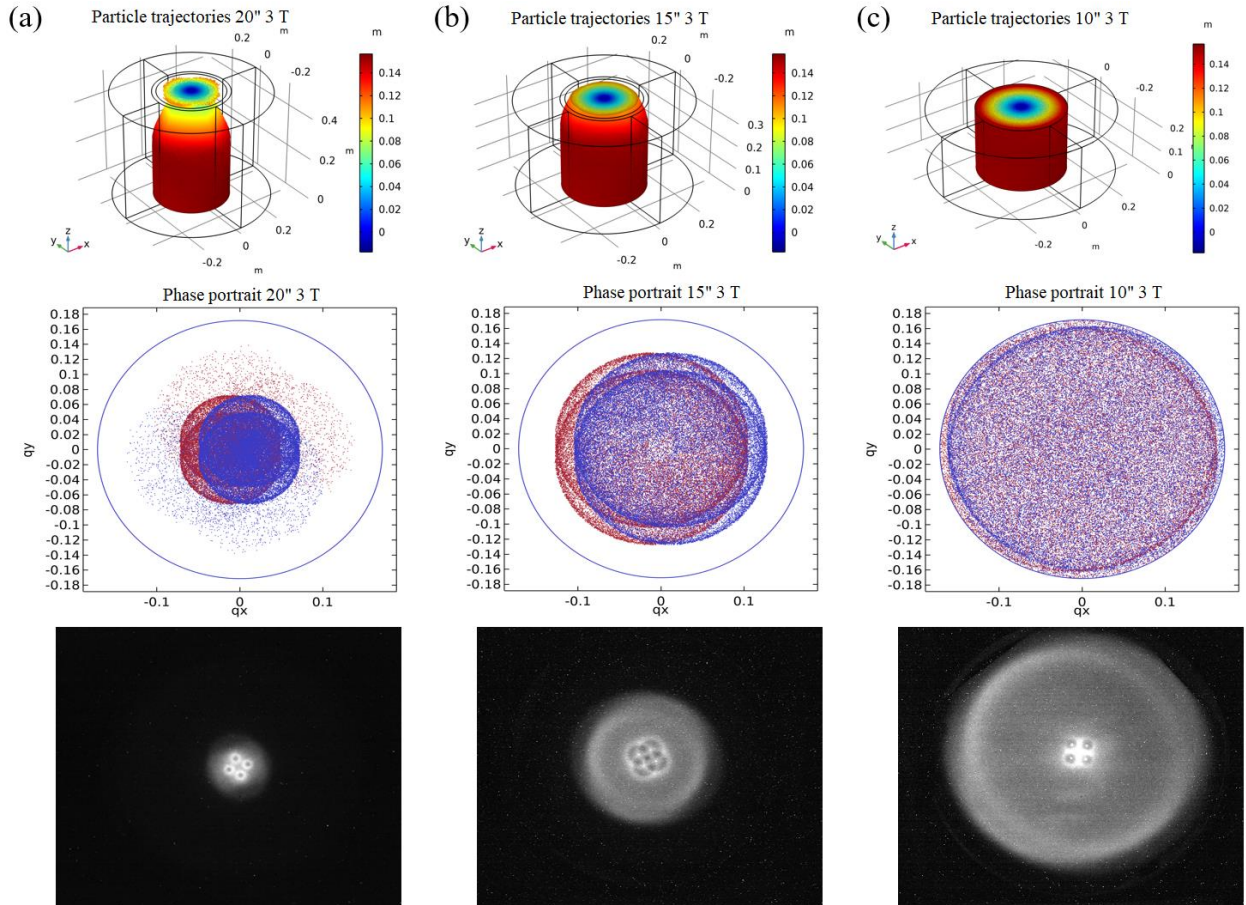

**Figure 5:** The particle trajectories within the vacuum cylinder at positions of 20", 15", and 10". *Top*: The colors indicate the radial displacement of each particle from its initial position. *Centre*: The Phase Portrait plot within COMSOL where the colors indicate the side of the plate the particle originated from. Respectively, red or blue particle tracks originated from the two sources on the left side or right side of the plate. *Bottom*: The experimental results showing patterns comparable to those observed in the simulated outcomes for the corresponding setups.

**Figure 6** compares the results derived from both the experiments and COMSOL simulations, illustrating the count of alpha particles reaching the scintillator plate emitted from the source plate containing $^{241}$Am, at varying distances. The experimental particle counts were estimated by analyzing images from the experiments using MATLAB image processing to calculate the pixel counts per second. In Figure 6, the solid blue squares represent the pixel counts per second of the image from the experiment at 0 T, while the blue crossed squares indicate the pixel counts per second at 3 T. Similarly, the solid red circles represent the simulated data for the number of particles reaching the detector at 3 T, and the crossed red circles show the simulated particle counts at 0 T. **Table 1** gives the detail summary of the analyzed data from both simulation and experiments.

**Table 1**. Data collected from the experiment and COMSOL simulation estimating number of particles reaching the scintillator plate.

| Source to Scintillator Distance | Experimental data Pixel Counts Per Sec | | Simulation data Number of Particles Reaching the Outlet/Detector | |
|---|---|---|---|---|
| (Inches) | (0 T) | (3 T) | (0 T) | (3 T) |
| 10 | $1.1\times10^6$ | $2.8\times10^6$ | $3.4\times10^4$ | $6.9\times10^4$ |
| 15 | | | $1.9\times10^4$ | $7.1\times10^4$ |
| 20 | $4.9\times10^5$ | $3.3\times10^6$ | $1.1\times10^4$ | $6.9\times10^4$ |

In the absence of a magnetic field (0 T), the escaping particles exhibit isotropic emission, with the count depending on the alpha particles that are emitted into a solid angle that corresponds to a clear line of sight between the sources and the scintillator plate. However, with a 3-T magnetic field, it is intriguing to observe a substantial increase in the count of escaping particles—6.5 times (~6.89 times from experiment) at a distance of 20", 4.0 times at 15 inches, and 2.0 times (~2.5 times from experiment) at 10 inches. Remarkably, the count of alpha particles reaching the scintillator under a 3-T magnetic field remains consistent. This consistency can be attributed to magnetic confinement, which reduces particle collisions with the chamber walls, thereby defining the thrust axis based on the Larmor motion of the particles. These findings not only indicate the potential of designing detector chambers with cooled CMOS cameras for detecting alpha particles under extreme conditions, such as high magnetic fields but also validate the feasibility of using high magnetic fields to constrain fission fragment trajectories. This approach holds promise for achieving high specific impulse and power density in advanced nuclear propulsion systems like the fission fragment rocket engine (FFRE), by employing a fissile fuel particle embedded in a high magnetic field.

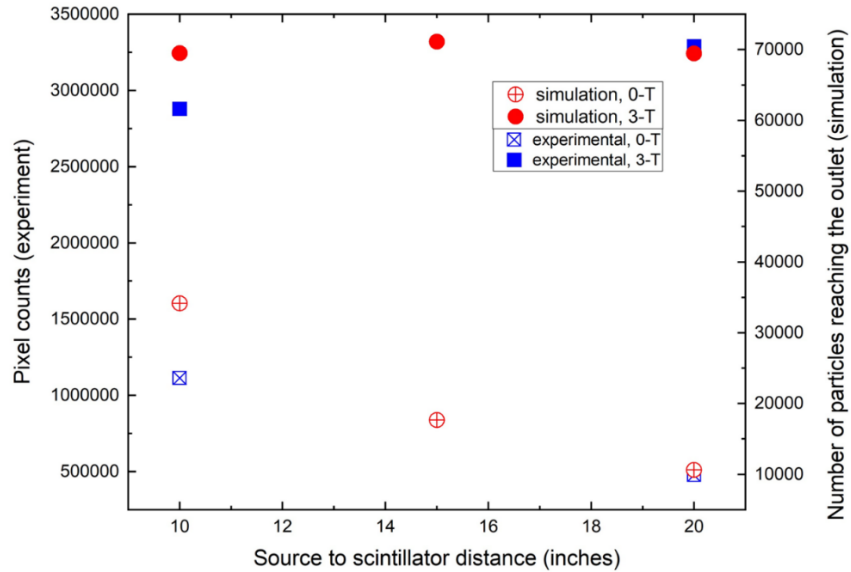

**Figure 6:** Comparison of experimental and simulation data for the count of alpha particles reaching the scintillator plate emitted from the source plate containing $^{241}$Am, at varying distances.

## 5.0 Conclusions

In this paper, an inventive system for detecting alpha particles was developed and built. The primary aim was to gain a deeper understanding of how thrust particles behave within the strong magnetic field of a 3-Tesla MRI system. Simulations were conducted employing COMSOL software and $^{241}$Am was chosen as the study material. The $^{241}$Am sources were then positioned within a cylindrical vacuum chamber and situated inside the superconducting magnet of a Siemens 3T MRI system. Through this configuration, the trajectories of emitted alpha particles were simulated and monitored.

With a 3-T magnetic field, it is intriguing to observe a substantial increase in the count of escaping particles—6.5 times at a distance of 20", 4.0 times at 15", and 2.0 times at 10" comparing to 0 T magnetic field. This observation indicates that a greater number of alpha particles strike the scintillator panel when exposed to the confinement of the magnetic field. These findings not only propose that the development of detection chambers equipped with cooled CMOS cameras holds potential for detecting alpha particles under extreme circumstances, including high magnetic fields, but they also verify the feasibility of utilizing powerful magnetic fields to control the trajectories of fission fragments. This strategy could potentially lead to achieving high specific impulse and power density in advanced nuclear propulsion systems, such as the fission fragment rocket engine (FFRE), through the utilization of fissile fuel particles embedded in high magnetic fields. Ongoing future experiments will involve the continuation of various approaches, including the exploration of higher magnetic HTS magnets.

## 6.0 Conflict of Interest

The authors declare that the research was conducted in the absence of any commercial or financial relationships that could be construed as a potential conflict of interest.

## 7.0 Author Contributions

All authors contributed to the conception and design of the research reported here. SP wrote the first draft of the manuscript. All authors contributed to manuscript revision, read, and approved the submitted version.

## 8.0 Funding

This work was supported by NIAC Phase I contract No. 80NSSC23K0592, the Texas Research Incentive Program, and 2025 Research Assistance Program by Texas Tech University.

## 9.0 Acknowledgments

The authors would like to thank Kasey Rieken, Brad Johnson, Gerald Jones for their useful discussions. The identification of commercial products, contractors, and suppliers within this article are for informational purposes only, and do not imply endorsement by Texas Tech University, their associates, or their collaborators.